\documentclass[12pt]{article}
\newcommand{\be}{\begin{equation}}
\newcommand{\ee}{\end{equation}}
\newcommand{\bqa}{\begin{eqnarray}}
\newcommand{\eqa}{\end{eqnarray}}

\begin{document}
\begin{center}
{\LARGE A short note on $q\bar qg$ hybrid assignment for $X(1812)\to \omega\phi$ }\\[0.8cm]
{\large Kuang-Ta Chao$^{(a,b)}$}\\[0.5cm]
{\footnotesize (a)~Department of Physics, Peking University,
 Beijing 100871, People's Republic of China}\\
{\footnotesize (b)~China Center of Advanced Science and Technology
(World Laboratory), Beijing 100080, People's Republic of China}
\end{center}
\vspace{0.5cm}

\begin{abstract}
BES Collaboration has recently observed an enhancement near the
$\omega\phi$ threshold in the double OZI suppressed decay $J/\psi
\to\gamma\omega\phi$. This $\omega\phi$ enhancement is probably
due to a new $J^{PC}=0^{++}$ X(1812) state. We discuss a possible
assignment that the X(1812) is  a $ 1/\sqrt{2}(u\bar u+d\bar d)g$
(or $s\bar sg$) hybrid meson. We suggest further measurements to
test these assignments.
\vspace{1cm}\\

\end{abstract}

\vspace{1cm} Based on a sample of $5.8\times 10^7 ~J/\psi$ events,
very recently the BES Collaboration has observed an enhancement
near threshold in the $\omega\phi$ invariant mass spectrum from
the double OZI suppressed decay of $J/\psi \to \gamma\omega\phi$
with a statistical significance of more than 10 $\sigma$. As a
hadronic resonance state $X$, this enhancement has the following
mass, width, and branching ratio product\cite{bes}: \bqa
  M=1812^{+19}_{-26}\pm {18}~MeV, ~~~~~~\Gamma=105\pm 20\pm 28~MeV,
   \\
  B(J/\psi \to \gamma X)\times B(X \to \omega\phi)=(2.61\pm 0.27\pm
  0.65)~10^{-4}.
\eqa Moreover, the spin-parity $J^{PC}=0^{++}$ is favored  for
$X(1812)$ by a partial wave analysis.

$J/\psi$ radiative decays into light hadrons proceed through
$J/\psi\to\gamma gg$ and are viewed as the glue-rich channels. The
striking feature of the X(1812) that it decays substantially to an
$\omega$, which is a $(1/\sqrt{2})(u\bar u+d\bar d)$ meson, and a
$\phi$, which is a $s\bar s$ meson, may imply that the X(1812) is
unlikely to be an ordinary $q\bar q$ meson, because the $q\bar q
\to q\bar q+q'\bar q'$ decays ($q$ and $q'$ are of different
flavors) should be suppressed, compared with the OZI allowed
$q\bar q \to q\bar q'+q'\bar q$ decays. Therefore, the X(1812) is
likely to be a new type of hadrons other than the conventional
$q\bar q$ mesons.

In the following we will discuss a possible assignment for the
X(1812). That is, the X(1812) is a $q\bar q g$ hybrid state.

The expected existence of hybrid states is an essential feature of
quantum chromodynamics. A hybrid state $q\bar q g$ is made of
$(q\bar q)_8$, a quark and an antiquark in color-octet, combined
with an excited gluon $g$. Hybrid states have been extensively
studied in various approaches such as the constituent gluon
model\cite{mandula}, the flux tube model\cite{flux}, the bag
model\cite{barnes,chanowitz}, the QCD sum rules\cite{sum}, and the
lattice QCD\cite{lattice,luo}. The experimental status for hybrid
states can be found in a mini-review for "Non-$q\bar q$
candidates" written by the Particle Data Group\cite{pdg}. As a
more fundamental theoretical approach for hybrid studies, lattice
QCD  has made interesting progress and shows that for an exotic
spin-parity $J^{PC}=1^{-+}$ hybrid, which does not mix with
ordinary $q\bar q$ mesons and is then a "pure" hybrid state, the
mass is estimated to be in the range 2.0(2)~GeV for $s$ quarks,
and 1.9(2)~GeV for $u,d$ quarks. The decay has also been studied
for the heavy quark hybrid, for which the dominant decay mode is
the string de-excitation into a flavor singlet meson, and the
magnitude of the decay width is found to be of order 100
MeV\cite{lattice}. Although the hadronic decays for light-quark
hybrid mesons has not been explored, the above lattice QCD results
do shed light on a rough estimate of decay widths and modes for
light-quark hybrids. Concerning the X(1812) observed by BES, it
seems that its mass, total width, and decay modes in particular,
might be compatible with a $0^{++}$ light-quark
$(1/\sqrt{2})(u\bar u+d\bar d)g$ hybrid state. In the flux tube
model the mass of $0^{++}$ hybrid is higher than that of $1^{-+}$
($1^{+-},1^{++},1^{--},0^{+-},0^{-+},2^{+-},2^{-+}$ as well), but
considering the exiting theoretical uncertainties including the
effects of mixing with other $0^{++}$ states, a mass of 1812 MeV
for the $0^{++}$ hybrid does not seem to be unacceptable.  At
present it is still premature to make precise quantitative
predictions for hybrid states due to our limited understanding of
nonperturbative dynamics. Therefore we will focus on some
distinctively qualitative features of the hybrid states and
discuss a possible hybrid meson interpretation for the X(1812).

As already noted in \cite{chanowitz,chao}, the decay into
$\omega\phi$ can be a distinctive feature of $0^{++}$ and $2^{++}$
$q\bar q g$ hybrid states including both  $(1/\sqrt{2})(u\bar
u+d\bar d)g$\cite{chanowitz} and $s\bar sg$ states\cite{chao}.
Since for the X(1812) the $\omega\phi$ mode is the only decay
channel observed so far, it is natural to propose that the X(1812)
be a possible candidate of the $0^{++}$ $q\bar q g$ states.

We first use a bag model estimation. In the spherical
approximation of the bag model, the $q\bar q$ can couple to either
the TE (transverse electric) or the TM (transverse magnetic)
gluon. The TE mode has an axial vector quantum number
$J^{PC}=1^{+-}$, while the TM mode has
$J^{PC}=1^{--}$\cite{chanowitz}. Moreover, in this model it was
found that the TE gluon couples in the s-channel to $u\bar u,
d\bar d,$ and $s\bar s$ in an approximately flavor symmetric way,
whereas the TM gluon couples to $s\bar s$ much stronger than to
$u\bar u$ and $d\bar d$\cite{chanowitz}. In this model, we may
assign X(1812) as the $0^{++}$ $(1/\sqrt{2})(u\bar u+d\bar
d)g_{TM}$  state, which has a mass in the range 1.51-1.90 GeV.
Note that in this model the gluon excitation $g_{TM}$ couples
dominantly to a color-octet $s\bar s$, so we would approximately
have \be g_{TM}\rightarrow (s\bar s)_8, \ee and expect this hybrid
to decay via \be X(1812)=\frac {1}{\sqrt{2}}(u\bar u+d\bar d)
g_{TM}\to \omega_8\phi_8\to \omega\phi, K^*\bar K^*, K\bar K,...
\ee Here the color-octet quark pairs $\omega_8$$\phi_8$ hadronize
into color-singlet meson pairs by gluon exchange into $\omega\phi$
(without spin flip), and $\eta\eta,\eta\eta'$ (with spin flip) or
quark rearrangement into $K^*\bar K^*, K\bar K$. The long-ranged
color-electric force can induce the non-spin-flip transition,
whereas short-ranged color-magnetic force causes the spin-flip
transition. If the long-ranged color-electric force is stronger
than the short-ranged color-magnetic force, we would expect the
$\omega\phi$ decay to be stronger than the $\eta\eta,\eta\eta'$
decays. Furthermore, if the hybrid decay proceeds mainly via
string de-excitation into a flavor singlet meson, as indicated by
the lattice calculation for the heavy quark hybrid meson (the
$1^{-+}$ charmonium hybrid $c\bar cg$ mainly dceays into
$\chi_{c1}+\sigma$ with $\sigma\to\pi\pi$)\cite{lattice}, we would
expect the $\omega\phi$ decay to be also stronger than the
$K^*\bar K^*, K\bar K$ decays. One of the distinctive features in
this model is that the decay into $\omega\omega$ will be much
weaker than into $\omega\phi$. Moreover, in this model, we would
predict the existence of the $s$ quark partner of the $u,d$ quark
hybrid $X'$, and its decay modes
 \be X'=s\bar
sg_{TM}\to\phi_8\phi_8\to\phi\phi,\eta\eta,\eta\eta',\eta'\eta',...
 \ee
Here the $X'$ has $J^{PC}=0^{++}$ and its mass may be roughly
estimated to be ranging from 2000 to 2100 MeV by taking the mass
difference between $s$ quark and $u,d$ quark to be 100-150 MeV.
However, if the mass of $X'$ is below the $\phi\phi$ threshold,
its $\phi\phi$ decay mode would not be observed. In any case, if
this model prediction makes sense, a weak signal of $0^{++}$
$X(1812)\to\omega\omega$, and a strong signal of $0^{++}$
$X'(2040-2100)\to\phi\phi$ should be observable experimentally. We
note that the MarIII Collaboration\cite{mark3} reported that the
$J/\psi\to\gamma\omega\omega$ decay has a branching ratio of
$(1.76\pm 0.09\pm 0.45)\times 10^{-3}$ and the $\omega\omega$
invariant mass distribution peaks at 1.8~GeV, but the
$\omega\omega$ system below 2~GeV is predominantly pseudoscalar.
Similar results were also reported by the DM2
Collaboration\cite{dm2}. It is therefore very desirable to check
the BES data for this channel to see if there is a $J^{PC}=0^{++}$
$\omega\omega$ component in the peak of 1.8~GeV.

We now turn into a more general discussion by assuming that the
gluon couples to the color-octet $u\bar u, d\bar d, s\bar s$
symmetrically in the decay of hybrid state $(1/\sqrt{2})(u\bar
u+d\bar d)g$
 \be g\rightarrow 1/\sqrt{3}(u\bar u+d\bar d+s\bar
s)_8.
 \ee
If this is the case, then the $X(1812)$ would have stronger decay
into $\omega\omega$ than $\phi\phi$, and stronger decay into
$\pi\pi$ than $K\bar K$. These predictions may also be tested by
experiment.

We might also interpret the $X(1812)$ as the $s\bar sg$ hybrid
meson. Then it would not decay into $\omega\omega$. However, its
mass is too small, compared with most predictions of theoretical
models and lattice QCD calculations.

\vspace{0.5cm} $Note~~ added. ~$ Related to the BES result
recently reported in \cite{bes}, a paper\cite{li} proposed a
four-quark state interpretation for the X(1812); and another
subsequent paper\cite{he} focused on the mixing between $0^{++}$
$q\bar q$ mesons, glueballs, and $q\bar qg$ hybrids, and used the
$f_0(1790)$=$ 1/\sqrt{2}(u\bar u+d\bar d)g$ as input and then
found $X(1812)$ to be mainly the $s\bar sg$ hybrid state.

\vspace{0.5cm}
$Acknowledgments.$~~The author would like to thank  X.Y. Shen for
valuable discussions on the feasibility for experimental test of
the proposed hybrid states. This work was supported in part by the
National Natural Science Foundation of China (No 10421503), the
Key Grant Project of Chinese Ministry of Education (No 305001),
and the Research Found for Doctorial Program of Higher Education
of China.

\end{document}